\def\year{2017}\relax
\documentclass[letterpaper]{article} 
\usepackage{aaai17}  
\usepackage{times}  
\usepackage{helvet}  
\usepackage{courier}  
\usepackage{url}  
\usepackage{graphicx}  
\usepackage{booktabs}
\usepackage{amsmath}
\usepackage{amssymb}
\usepackage{xcolor}

\newcounter{NumberOfComments}
\stepcounter{NumberOfComments}

\newcounter{JSNumberOfComments}
\stepcounter{JSNumberOfComments}

\begin{document}
%
\title{Data-driven Approach to Measuring the Level of Press Freedom Using \\ Media Attention Diversity from Unfiltered News}
\author{Jisun An and Haewoon Kwak\\
Qatar Computing Research Institute\\
Hamad Bin Khalifa University\\
Doha, Qatar\\
\{jisun.an, haewoon\}@acm.org
}
\maketitle
\begin{abstract}
Published by Reporters Without Borders every year, the Press Freedom Index (PFI) reflects the fear and tension in the newsroom pushed by the government and private sectors. While the PFI is invaluable in monitoring media environments worldwide, the current survey-based method has inherent limitations to updates in terms of cost and time.  In this work, we introduce an alternative way to measure the level of press freedom using \textit{media attention diversity} compiled from Unfiltered News. 
\end{abstract}

\section{Introduction}

Media that is independent, providing reliable news and information and generating debate and pluralism, is a key to democracy. 
Having a measure of such media environment, thus, can benefit the societies--it can be used as a benchmark to advocate for greater press freedom to governments. 
One of such efforts is the development of Press Freedom Index (PFI) by Reporters Without Borders (RSF). Every year since 2002, RSF has published the PFI, which ranks 180 countries by the level of freedom available to journalists~\cite{pressfreedom}. To calculate the Index, RSF uses the responses of experts to a questionnaire and a manually constructed data on abuses by in-house specialists. 

These acts of measuring the level of press freedom have been facing a challenge in adapting their methodology to a rapidly changing news and media environment. 
Monitoring press freedom levels would be beneficial to those in need of attention.
Especially in an era when the global media freedom is decreasing mainly due to the increase of authoritarian tendencies of governments and foreign attacks\footnote{https://rsf.org/en/deep-and-disturbing-decline-media-freedom}, such ability would be helpful.
However, the methodology to compile PFI requires a lot of effort, time, and money. 

To overcome such limitation, in this work, we introduce a data-driven approach to measure the level of press freedom using \textit{media attention diversity} computed from large-scale online news data. 
To this end, we collect data from Unfiltered News, which is a visualization tool that allows people to explore global news, and compute media attention diversity. 
We firstly examine media attention diversity across countries and then validate that this media attention diversity has a stronger explanatory power for PFI than any major  country-level attributes.

\section{Background}
\subsection{Press Freedom Index}

Published by Reporters Without Borders (RSF) since 2002, the World Press Freedom Index (PFI) has been widely used as a point of reference by global media. 
PFI ranks 180 countries by the level of freedom available to journalists. In other words, it is indicative of a ``climate of fear and tension combined with increasing control over newsrooms by governments and private-sector interests.'' Diplomats and international entities such as the United Nations and the World Bank are also using it for their work.

PFI is compiled by 1) the data collected through a questionnaire, which includes 87 questions regarding pluralism, media independence, environment and self-censorship, legislative framework, transparency, and infrastructure and 2) the data on abuses, which is a detailed tally on abuses and violence against journalists and media outlets in different regions collected by in-house specialists.  
The lower value PFI is, the better press freedom the country has.  For example, the PFI of Finland is 8.59 (1st), and that of North Korea is 83.76 (179th) as of 2016.   

Previous studies have examined what other indicators are related to PFI and found it is associated with the degree of development, the level of poverty, and the governance~\cite{guseva2008press}. More recently, Asal and Hoffman~\citeyear{asal2016media} studied the relationship between PFI and foreign attacks and found they are not related. Our work can complement this work by providing a data-driven approach to measuring the level of press freedom using online news data.

\subsection{Unfiltered News}

Unfiltered News\footnote{http://unfiltered.news/} was developed by Google Ideas, now Jigsaw, to address the problem of the filter bubble by easily browsing all the online news worldwide. 
Incorporating a virtue of big data and advanced language translation techniques to process all the news data available in Google News, Unfiltered News enables users 1) to find news beyond their border--topics not necessarily popular within their regions, 2) to see  which regions are reporting on a particular topic, and 3) to read topics and headlines in any language. 

\begin{figure} [hbt!]
  \begin{center}
  \includegraphics[width=\columnwidth]{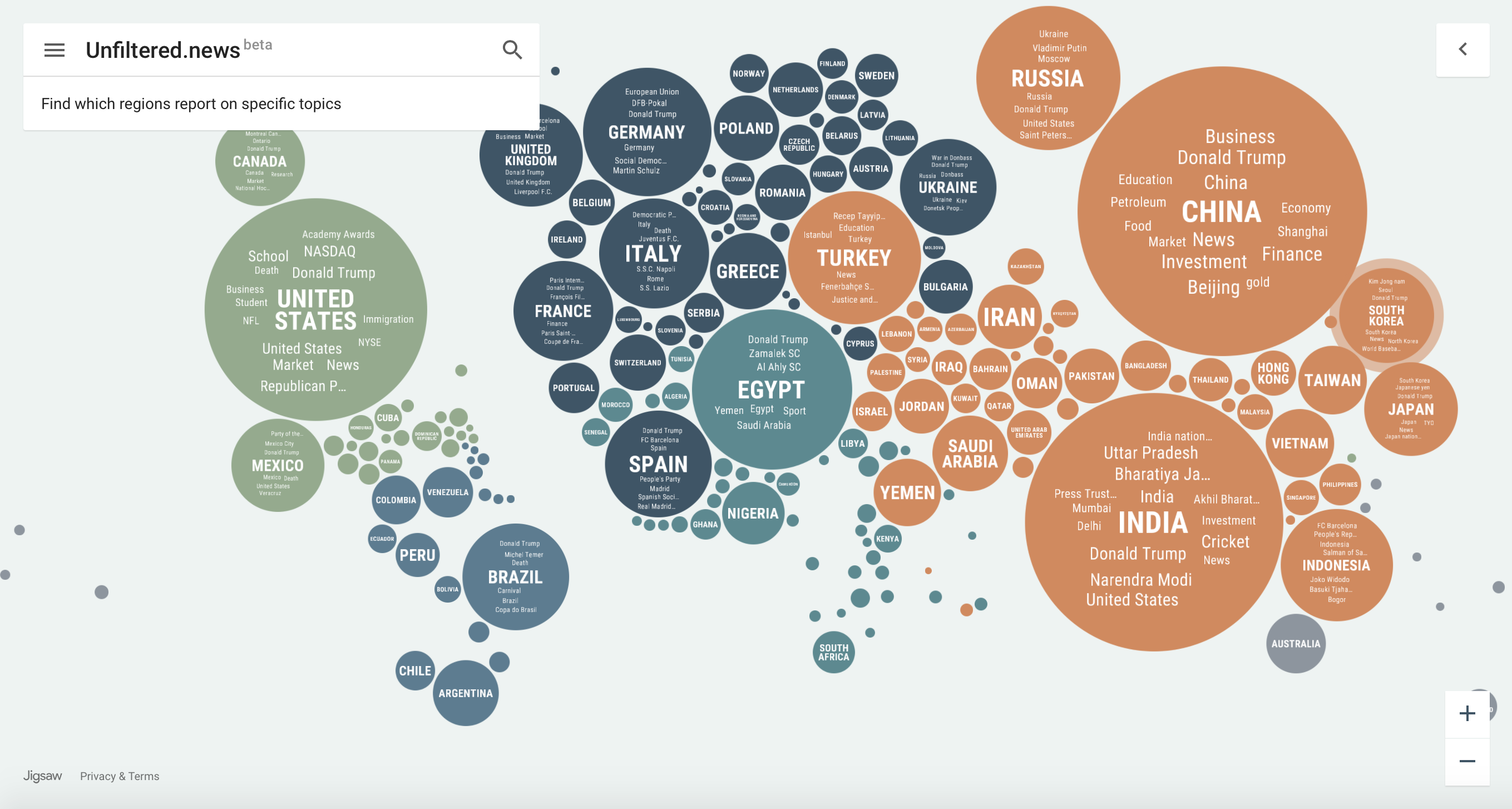}
  \caption{Screenshot of Unfiltered News}  
  \label{fig:screenshot}
  \end{center}
\end{figure}

Figure~\ref{fig:screenshot} presents the screenshot of the top page of Unfiltered News.
Users can explore a list of the most popular topics\footnote{Unfiltered News uses an entity defined in Google Knowledge Graph as a unit of ``topic.''} in each country by zoom-in and -out.

\section{Data Collection}

Unfiltered News offers two kinds of indexed data for each country: the top 100 most frequently mentioned topics and the 100 topics that are least mentioned (compared to those in other countries). In this work, we focus only on the former, the most frequently mentioned topics in each country. 
In other words, we examine what news media in a country pays attention to.

We collect the most popular topics in 196 countries that are available in Unfiltered News from 7 March to 9 October 2016.  
We develop crawlers with reasonable inter-request times so that the performance of the server does not degrade.

\subsection{Data Description and Notation}

Although Unfiltered News ideally provides the 100 most popular topics for each country every day, sometimes it returns less than 100 topics for a certain country on a particular day. Prior to the analysis, we eliminate such incomplete data.  

We denote by $\mathrm{C}=\{c_1, c_2, ..., c_n\}$ a set of $n$ countries and $\mathrm{D}=\{d_1$, $d_2$, ..., $d_m\}$ a set of $m$ days available in the data collection. 
We denote by $\mathrm{M}^{c_i}_{d_j}(k)$ the $k$ most popular topics in the country $c_i$ on the day $d_j$. For example, {\small$\mathrm{M}^{Korea}_{20160101}(10)$} is the top 10 most frequently mentioned topics in Korea on 1 January 2016.
We then eliminate incomplete data in the following ways:

\begin{enumerate}
    \item We set the threshold parameter $k$ (0 $<$ k $\leq$ 100). 
    \item For each country $c_i$$\in$$\mathrm{C}$, we put $c_i$ in $\mathrm{C}^k$ only when $c_i$ has at least the top $k$ topics on $\forall d_i$$\in$$\mathrm{D}$. In other words, we eliminate all the countries that have less than the top $k$  topics even on a single day.  
    \item Let us say $\mathrm{C}^k$=$\{c^k_1$, $c^k_2$, ..., $c^k_l\}$. Again, $|\mathrm{M}^{c^k_i}_{d_j}(k)|$ $\geq$ $k$ for any $1 \leq i \leq l=|\mathrm{C^k}|$ and $1 \leq j \leq m=|\mathrm{D}|$. 
    \item Then, the final dataset of the top $k$ topics, $\mathrm{M}(k)$ is: 
    {\small \frenchspacing
    \\ $\mathrm{M}(k)$=$\{(\mathrm{M}^{c^k_1}_{d_1}(k)$,$\mathrm{M}^{c^k_1}_{d_2}(k)$,...,$\mathrm{M}^{c^k_1}_{d_m}(k))$, $(\mathrm{M}^{c^k_2}_{d_1}(k)$, $ \mathrm{M}^{c^k_2}_{d_2}(k)$, ..., $\mathrm{M}^{c^k_2}_{d_m}(k))$, ..., $(\mathrm{M}^{c^k_l}_{d_1}(k)$, $ \mathrm{M}^{c^k_l}_{d_2}(k)$, ..., $\mathrm{M}^{c^k_l}_{d_m}(k))\}$
    }
\end{enumerate}

In summary, after preprocessing the data, we get the data that consists of countries that have the top $k$ topics every day. For example, when $k=100$, the resulting data includes countries that have at least the top 100 topics every day from 7 March to 9 October 2016.

If we do not have any constraint ($k$=0), the number of countries in $\mathrm{C}^0$ is 196.  
With a weak condition ($k$=10), $|\mathrm{C}^{10}|$ quickly decreases to 129.  
The number of countries in $\mathrm{C}^k$ monotonically decreases with growing $k$ and reaches at 88 when $k$=90. This means that these 88 countries have at least the 90 most popular topics for every single day during our data collection period.  
We note that no countries have the top 100 topics ($|\mathrm{C}^{100}|=0$) every day.

\section{Diversity of Media Attention}

In this section, we first show that the diversity of media attention is considerably different across countries. Also, by examining which country has more (or less) diverse media attention, we can get an insight that media attention diversity might be related to media environments.

\subsection{Measures of Media Attention Diversity}

We define the media attention diversity of a given country by the cardinality of a set of top topics in the country during a certain period. 
More formally, $\mathrm{U}^{c_j}(k) = |  \bigcup\limits_{d_i}^{\mathrm{D}} \mathrm{M}^{c_j}_{d_i}(k)|$ is  the media diversity of country $c_j$ during $\mathrm{D}$. In the rest of the paper, we use the same $\mathrm{D}$, which is the entire data collection period from 7 March to 9 October 2016, and thus omit it for clarity. 
The bigger the $\mathrm{U}^{c_j}(k)$ is, the more diverse media attention of the country $c_j$ is.

\subsection{Media Attention Diversity across Countries}

\begin{figure} [hbt!]
  \begin{center}
  \includegraphics[width=40mm]{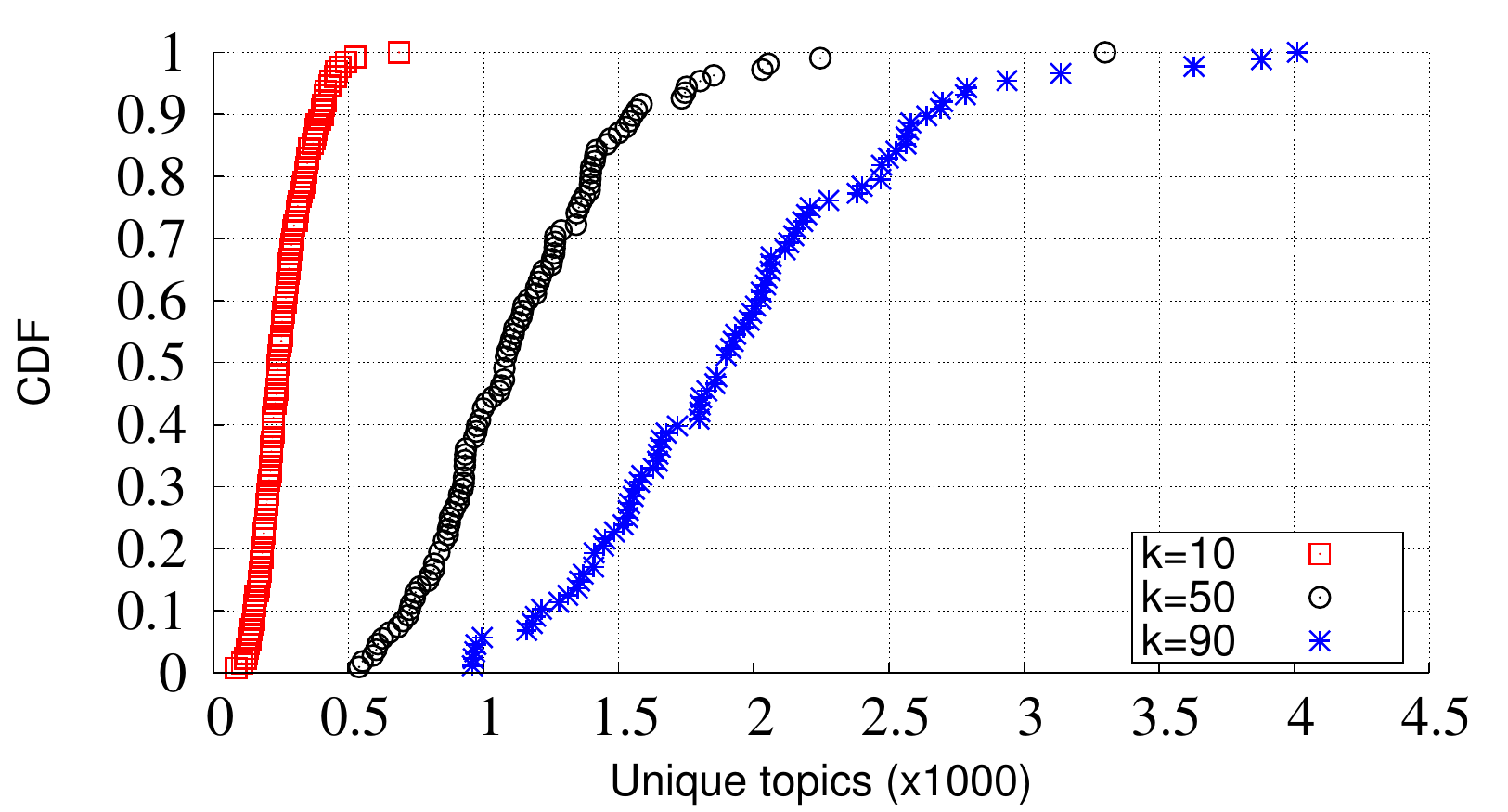}
  \includegraphics[width=40mm]{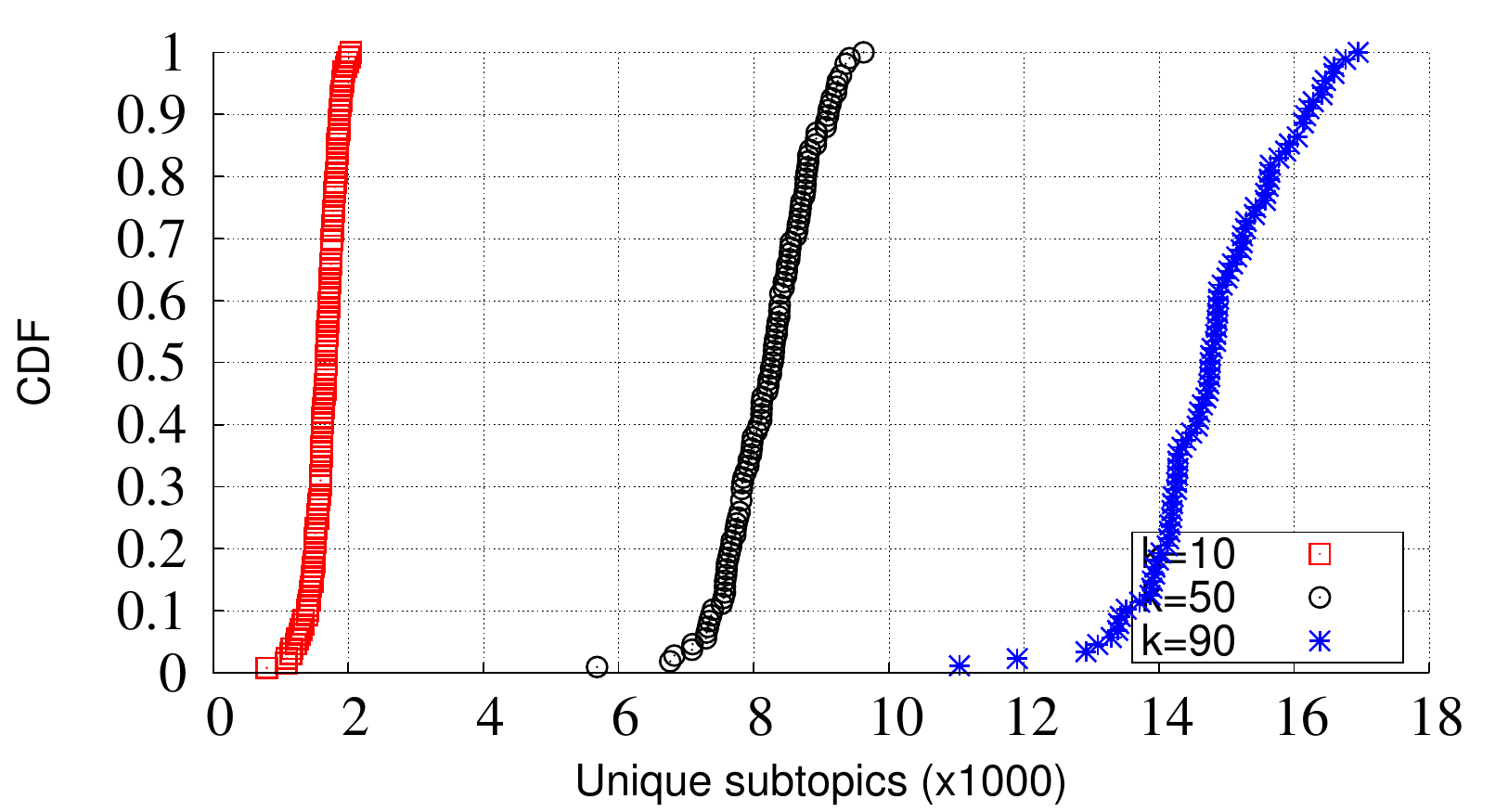}  
  \caption{Diversity of media attention across the countries in (a) topic-level (left); (b) subtopic-level (right)}  
  \label{fig:n_topics_cdf}
  \end{center}
\end{figure}

We find stark differences in the diversity of the media attention across the countries.  
For example, when $k$=90, we find that $\mathrm{U}^{\text{Luxembourg}}(90)= 4,012$ (the highest) and $\mathrm{U}^{\text{Egypt}}(90)= 959$ (the lowest).  In other words, in the course of 211 days, if we collect the top 90 frequently mentioned topics by Egyptian news media every day, the resulting set contains only 959 unique topics. It means that the popular topics are highly overlapped day by day in Egypt.
Compared to Luxembourg, the media attention diversity in Egypt is highly limited.  Figure~\ref{fig:n_topics_cdf} (a) shows the distribution of the diversity of media attention when $k$=10, 50, and 90.  

By definition, when $k$ increases, the diversity of media attention is likely to increase.   
It is, however, unexpected how limited the media attention of each country is.
In the left-most distribution represented by the red squares ($k$=10), the minimum is only 86 (Yemen). A union of the top 10 topics drawn from each of the 211 days contains only 86 unique topics for Yemen.  
Along with Yemen, other countries in the Middle East, Iraq (107), Saudi Arabia (120), Egypt (124), and United Arab Emirates (126), are the bottom five countries in terms of the least media attention diversity.  
As we mentioned above, even when $k$=90, Egypt has highly limited media attention diversity, which is only $\mathrm{U}^{\text{Egypt}}(90)= 959$.

The low diversity of media attention might be an artificial effect of the coarse granularity of topics, such that Barack Obama is counted as a single topic even though he is mentioned multiple times for different reasons. 
To relieve the limitations of coarse granularity, we use additional information Unfiltered News offers, called co-mentions, which are subtopics clarifying a given topic. 
For example, news mentioning Barack Obama can be about health care, Michelle Obama, or his public speech. In this case, health care, Michelle Obama, or his public speech are co-mentions (subtopic), respectively.  Thus, co-mentions with a given topic clarify the context of the topic in finer granularity.   
We use $l$ co-mentions with the original topic to define the subtopic.

Figure~\ref{fig:n_topics_cdf}(b) shows media attention diversity represented by subtopics ($l$=3). 
While the cardinality of a union set of subtopics for each country becomes larger than that of when using topics, the Middle East countries, such as Yemen and Egypt, which have less diversity at the topic level, still have less diversity at the subtopic level (e.g. $\mathrm{U}^{\text{Yemen}}_{\text{subtopic}}(10)$=795 (the lowest)). 
By manual inspection, these countries are likely to focus on domestic issues and regional (the Middle East) issues only, and thus the media attention diversity is limited. 

Next, we build a hierarchical regression model to use media attention diversity of a country to explain the level of the press freedom of the corresponding country.

\section{Measure the Level of Press Freedom}

\subsection{Correlation between Media Attention Diversity and Press Freedom Index}

The PFI reflects the degree of freedom available to journalists, news organizations, and netizens. While many factors can influence the PFI of a country, we begin with examining the impact of the media attention diversity. We later consider all other indicators when building a model.

\begin{figure} [hbt!]
  \begin{center}
  \includegraphics[width=75mm]{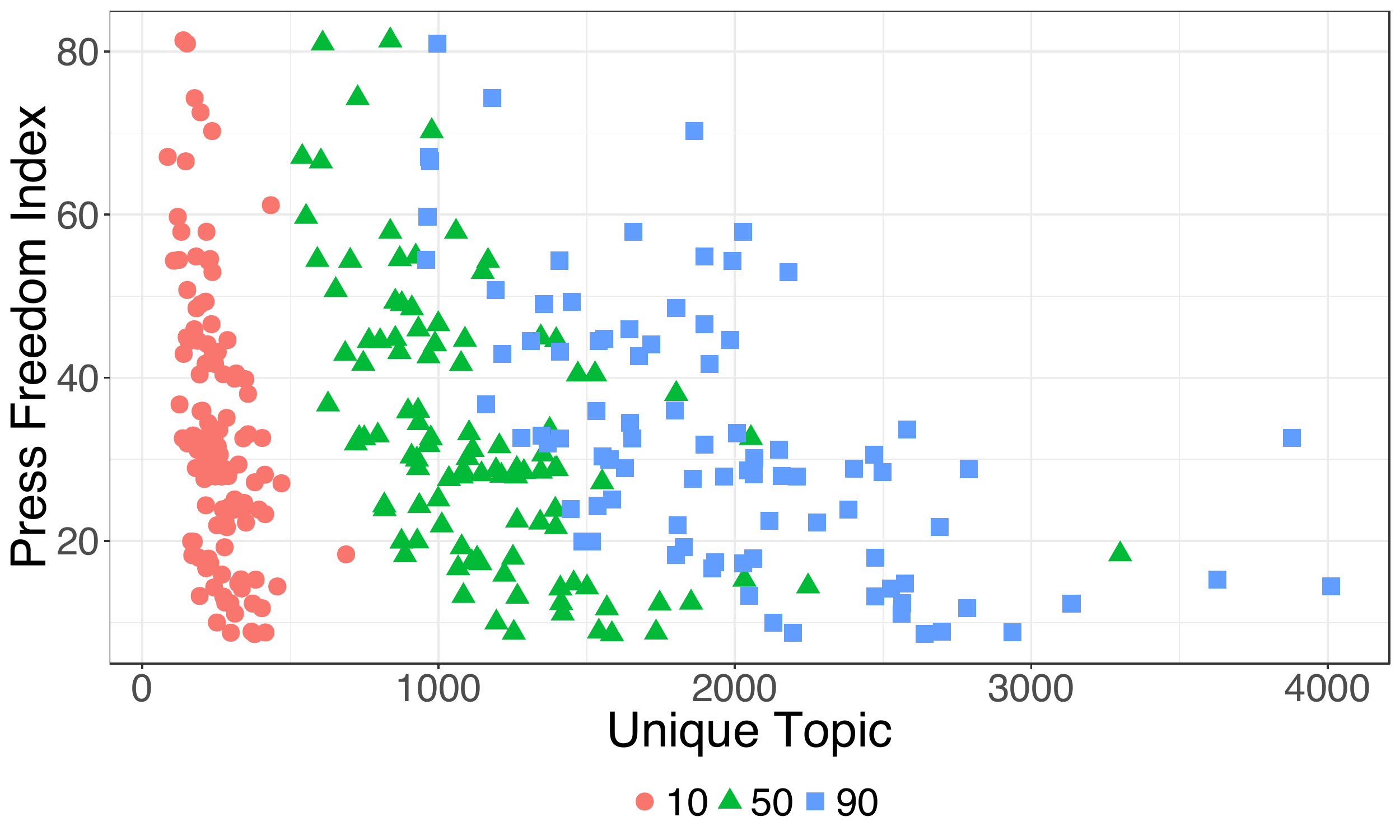}
  \caption{Correlation between $|\mathrm{U}^{c_i}(k)|$ and the press freedom index of country $c_i$ with varying $k$}  
  \label{fig:topic_freedom_with_different_k}
  \end{center}
\end{figure}

\begin{figure} [hbt!]
  \begin{center}
  \includegraphics[width=75mm]{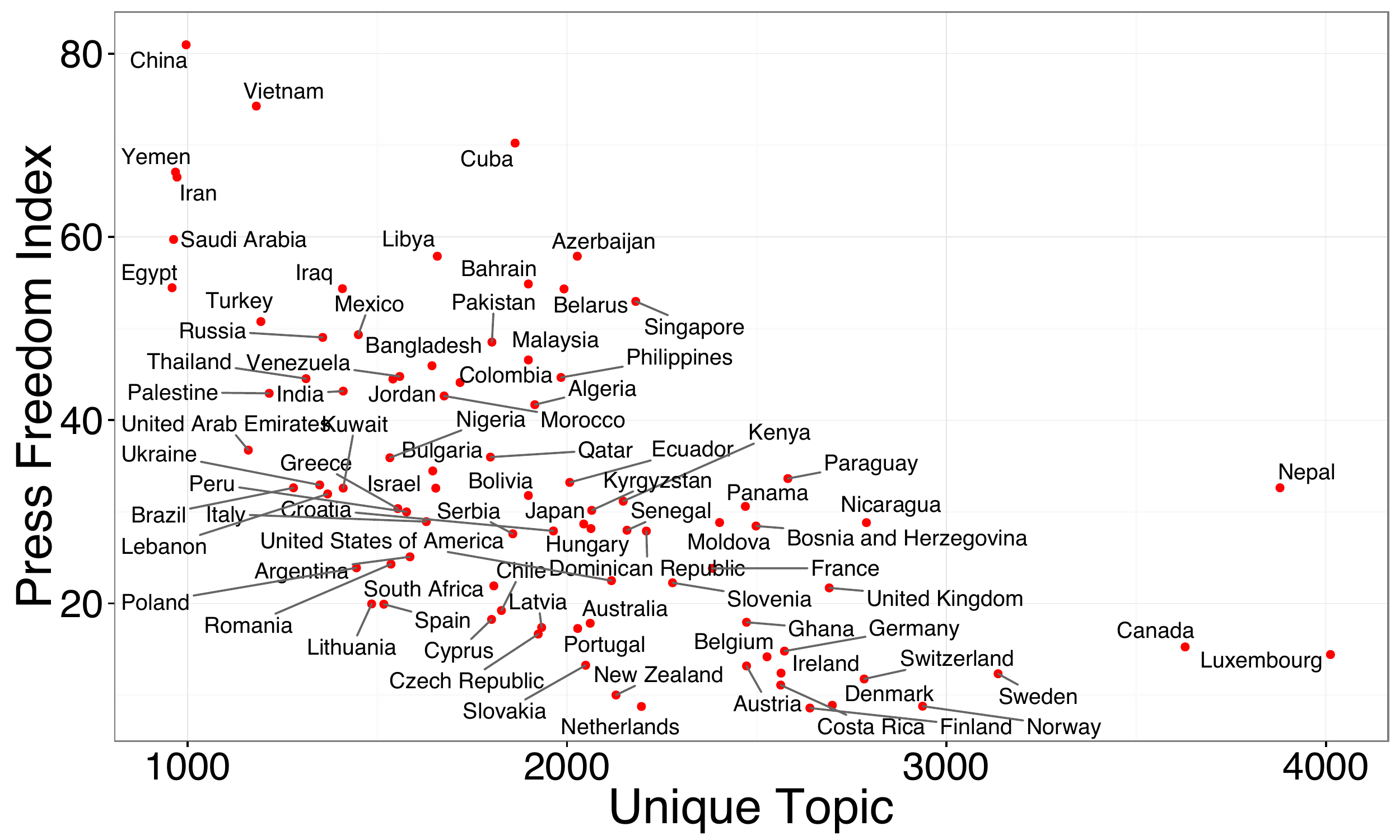}
  \caption{Correlation between $\mathrm{U}^{c_i}(90)$ and the press freedom index of country $c_i$}  
  \label{fig:topic_freedom}
  \end{center}
\end{figure}

Figure~\ref{fig:topic_freedom_with_different_k} presents how the relationship between media attention diversity of one country and its PFI changes with varying $k$. From the figure, we observe the strongest negative trend when $k$=90. The trend is statistically significant with the Pearson correlation coefficient, $r$, of -0.599 (95\% CI=[-0.718, -0.446]) when $k$=90. The negative correlation coefficient is consistently found for different $k$: $r$=-0.484 (95\% CI=[-0.608, -0.336]) when $k$=10, and $r$=-0.529 (95\% CI=[-0.653, -0.377]) when $k$=50.  
This negative correlation means that the country with a less diverse pool of popular topics is likely to have the worse press freedom (the larger PFI values).

Figure~\ref{fig:topic_freedom} shows the relationship between the media attention diversity and the PFI when $k$= 90 with the country labels. We note that, in the figure, we eliminate eight countries from $\mathrm{C}^{90}$ whose press freedom index is not available. 
The strong negative trend gives us an insight that the estimation of PFI by using media attention diversity would work.

\subsection{Hierarchical Regression Model}

We now build a hierarchical (mixed-effect) multiple regression model whose dependent variable is press freedom index, and the independent variables are media attention diversity and additional tens of national attributes~\cite{worldbank} of the country $c_i$. 
We choose the regression among various statistical models because interpreting the results is straightforward, while other models might improve the performance of the prediction task. However, in this work, our focus is not to fine-tune the result but show the potential of the data-driven way to assess the level of press freedom. 
We choose the hierarchical regression model to control a random effect driven by country-level variations.  
After checking the availability, multicollinearity, and distribution of national attributes, we finally include five variables in the model: log(media attention diversity), mobile cellular subscriptions (per 100 people), log(GDP per capita), log(population), and unemployment rate (\% of total labor force).  We confirm no collinearity among the five variables by a VIF test on the resulting model; all the remaining variables have VIF below than 1.6.  

\begin{table}
\begin{center}
\small \frenchspacing
\begin{tabular}{l c c c}
\toprule
                & Model 1 & Model 2 & Model 3 \\
\hline
(Intercept)     & $297.18^{***}$ & $29.13$       & $328.40^{***}$ \\
                & $(33.84)$      & $(26.80)$     & $(49.09)$      \\
log(attention diversity)      & $-35.08^{***}$ &               & $-32.81^{***}$ \\
                & $(4.49)$       &               & $(4.84)$       \\
cellular        &                & $0.05$        & $-0.05$        \\
                &                & $(0.06)$      & $(0.05)$       \\
log(GDP per capita)      &                & $-5.51^{***}$ & $-3.79^{**}$   \\
                &                & $(1.38)$      & $(1.14)$       \\
log(Population) &                & $3.04^{**}$   & $-0.16$        \\
                &                & $(1.14)$      & $(1.03)$       \\
unemployment    &                & $-0.16$       & $-0.54^{*}$    \\
                &                & $(0.28)$      & $(0.23)$       \\

\midrule
marginal $\textrm{R}^2$ & 0.4181 & 0.2842 & 0.5367 \\
 (fixed effects) & & & \\
conditional $\textrm{R}^2$ & 0.9283 &  0.9117 &  0.9429 \\
(random effects) & & & \\
AIC             & 680.90         & 709.05        & 669.29         \\
BIC             & 690.62         & 725.81        & 688.34         \\
Log Likelihood  & -336.45        & -347.52       & -326.64        \\
\bottomrule
\multicolumn{4}{r}{\scriptsize{$^{***}p<0.001$, $^{**}p<0.01$, $^*p<0.05$}}
\end{tabular}
\caption{Hierarchical regression analysis between media attention diversity and press freedom index.}
\label{table:coefficients}
\end{center}
\end{table}

Table~\ref{table:coefficients} presents the regression results for the press freedom index. We build three models, which are Model 1 incorporating media attention diversity only, Model 2 with all the national attributes except media attention diversity, and Model 3 including all the variables.   
As our models are mixed effect models, we report both $\mathrm{R}^2$ based on the fixed effects only (marginal $\mathrm{R}^2$) and based on the random effects as well (conditional $\mathrm{R}^2$).  

In Model 1, we can see the log-transformed media attention diversity is statistically significant.  The coefficient is -35.08, which means that a one percent increase in media attention diversity is associated with a (-35.08/100) unit decrease in press freedom index.  Again, a low press freedom index reflects a better media environment.  

In Model 2, we can see that two variables, GDP per capita and population, are statistically significant.  The higher GDP per capita is, the better the press freedom index.  In terms of population, the higher the population, the worse the press freedom index. The main reason for this conclusion is that some countries with a huge population, like China, India, or Egypt, have a low press freedom index, while European countries with a lower population have a high press freedom index, as in Figure~\ref{fig:topic_freedom}. In Model 3, which includes all the variables, we find that media attention diversity, GDP per capita, and unemployment are statistically significant. Population, which is significant in Model 2, becomes insignificant in Model 3.  

One unexpected finding is that, in Model 3, even though it is marginal, the level of unemployment has a negative relationship with press freedom index. From our manual inspection, we observe that countries with low unemployment rates do not necessarily have better press freedom. For example, the unemployment rate of Vietnam is 2.3\%, but its press freedom index is 74.3\%. While the unemployment rate itself does not have a meaningful explanatory power for PFI, after controlling other variables, it becomes significant.

Examining $\mathrm{R}^2$ values, Model 1 with a single variable, media attention diversity, has higher explanatory power than Model 2 with remaining national attributes. By using all variables, Model 3 explains 53.67\% of the data.


\section{Conclusion and Future Work}
In this work, we propose a data-driven approach to measure the level of press freedom using  \textit{media attention diversity}. 
Using large-scale news data collected from Unfiltered News, we build a hierarchical regression model to explain the press freedom index using several national attributes and media attention diversity. We find that media attention diversity itself has a strong explanatory power for press freedom index as compared to other national attributes. 

While our work proves news data can potentially be used to assess the level of press freedom, there is still room for improvement.  
We expect that media attention toward political topics might better reflect the level of press freedom rather than media attention across all topics. 
Also, we will examine whether using subtopics improves the model. 
The use of other statistical models instead of a hierarchical regression could improve the performance of the estimation of PFI.  
Finally, as we mentioned earlier, one advantage our method brings is that the level of press freedom can be instantly computed with news data rather than through a time-consuming survey, allowing us to track the media attention diversity continuously and detect the point at which it dramatically changes.

\small
\bibliographystyle{aaai}
\bibliography{2017.03-neco17-topic-diversity}  

\end{document}